\documentclass{amsart}

\usepackage{latexsym}
\usepackage{amsmath}
\usepackage{amsthm}
\usepackage{amssymb}
\usepackage{mathrsfs}
\usepackage{wasysym,color}
\usepackage{graphicx}

\usepackage{amsfonts}

\usepackage{graphicx}%
\newtheorem{theorem}{\indent Theorem}[section]
\newtheorem{proposition}[theorem]{\indent Proposition}

\newtheorem{lemma}[theorem]{\indent Lemma}
\newtheorem{remark}[theorem]{\indent Remark}
\newtheorem{corollary}[theorem]{\indent Corollary}
\newtheorem{observation}[theorem]{\indent Observation}

\numberwithin{equation}{section}

\newcommand{\R}{\mathbb{R}}
\newcommand{\eu}{u^s}
\newcommand{\au}{u^\alpha}
\newcommand{\ue}{u^\epsilon} 

\title{New look at the Navier-Stokes equation}    
\author{Tomasz Dlotko}\thanks{Tomasz Dlotko, Institute of Mathematics, Silesian University, Poland, tdlotko@math.us.edu.pl, Telephone: (+48)322582976}
\address[Tomasz Dlotko]{Institute of Mathematics, Silesian University, 40-007 Katowice, Bankowa 14, Poland} 
\email{tdlotko@math.us.edu.pl}

\subjclass[2000]{35Q30, 76D05, 35S15, 35A25.}  
\begin{document}
\date{}
\begin{abstract}
We propose a new way of looking at the Navier-Stokes equation (N-S) in dimensions two and three. In 2-D that problem is {\it critical} with respect to the standard $L^2$ a priori estimates. We consider its regular approximations in which the $-P\Delta$ operator is replaced with the fractional power $(-P\Delta)^{1+\alpha}, \alpha>0$ small, where $P$ is the projector on the space of divergence-free functions. The 3-D N-S equation is {\it super-critical} with respect to the standard $L^2$ a priori estimates; the regular approximating problem in 3-D should contain fractional power $(-P\Delta)^s$ with $s>\frac{5}{4}$. 

Using Dan Henry's {\it semigroup approach} we construct regular solutions to such approximations. The solutions are unique, smooth and regularized through the equation in time. Solution to 2-D and 3-D N-S equations are obtained next as a limit of the regular solutions of the above approximations. Moreover, since the nonlinearity of the N-S equation is of {\it quadratic type}, the solutions corresponding to small initial data and small $f$ are shown to be global in time and regular. 
\end{abstract}
\keywords{3-D Navier-Stokes equation; solvability; a priori estimates; fractional approximations}  
\maketitle
\thanks{{\large Dedicated to my Professors: Andrzej Lasota$^\dagger$, Jack K. Hale$^\dagger$ and Nick D. Alikakos.}}

\section{Introduction} 
Fifteen years ago, in the monograph \cite{C-D}, we were studying a direct generalization of {\it semilinear parabolic equations}, namely abstract semilinear equations with sectorial positive operator in the main part. Following the idea of Dan Henry \cite{HE}, such equations will be studied using classical techniques of the theory of ordinary differential equations, modified however to cover equations with unbounded operator in a Banach space. Such approach, located inside the semigroup theory, proves its strength and utility in the study of several classical problems; some of them were reported in \cite{HE, C-D}. But such technique offers further possible generalizations, first to study the problems, like e.g. Korteweg-de Vries equation and its extensions \cite{DL, D-S, D-K-Sh, D-K-Y}, where the solutions are obtained as a limit of solutions to {\it parabolic regularizations} of such equations (the method known as {\it vanishing viscosity technique}, originated by E. Hopf, O.A. Oleinik, P.D. Lax in 1950th). Another possible application of Henry's technique is to study {\it critical problems} (e.g. \cite{WA, C-D}), not falling directly into the class of semilinear sectorial equations because the nonlinear term in it is {\it of the same order, or value}, as the main part operator. Our recent paper \cite{D-K-S1} was devoted to such type problem, the {\it quasi-geostrophic equation in $\mathbb{R}^2$} (e.g. \cite{C-C-W, C-C, C-C1, K-N-V, W1, W2}). Another paper \cite{D-K} was devoted to {\it fractional Hamilton-Jacobi equation} in sub-critical case. But certainly the most celebrated example of such critical problem is {\it the Navier-Stokes equation in dimension two}. We will obtain and study its solutions constructed as limits of solutions to {\it sub-critical approximations} when $\alpha \to 0^+$, where the $P\Delta$ operator is replaced with its fractional power $-(-P\Delta)^{1+\alpha}, \alpha>0$ ($P$ is the projector on the space of divergence-free functions; see e.g. \cite{F-T}). The 3-D N-S equation is {\it super-critical} in a sense specified in subsection \ref{supercrit}. A higher order diffusion term like $-(-P\Delta)^s, s> \frac{5}{4}$, is needed in 3-D to guarantee global in time existence and regularity of solutions to the corresponding regularization of the N-S equation.

The classical $3-D$ Navier-Stokes equation considered here has the form:
\begin{equation}\label{NS-equation}
\begin{split}
&u_t = \nu \Delta u - \nabla p - (u \cdot\nabla)u + f,   \quad div u = 0, \quad x \in \Omega,\, t>0,  \\
&u = 0, \quad t>0, \quad x \in \partial \Omega,   \\ 
&u(0,x) = u_0(x),
\end{split}
\end{equation}
where $\nu>0$ is the viscosity coefficient, $u = (u_1(t,x), u_2(t,x), u_3(t,x))$ denotes velocity, $p = p(t,x)$ pressure, and $f = (f_1(x), f_2(x), f_3(x))$ external force, and $\Omega$ is a bounded domain with $C^2$ boundary. It is impossible to recall even the most important results devoted to that problem, since the corresponding literature is too large; see anyway \cite{F-H-T, F-T, GI, G-M, L, TE1, TE2, WA, C-D1} together with the references cited there.       

In space dimensions 2 and 3 the N-S equation possess local in time regular solutions, as stated in Theorems \ref{localexistence} and \ref{localexistence'}. We analyze further {\it criticality} of the Navier-Stokes equation (compare \cite{WA, C-D}), in a sense that available for it $L^2(\Omega)$ a priori estimate \eqref{1.9} is not sufficient to control its nonlinearity through the viscosity term $\nu P\Delta u$. Consequently, possible is a balance between the income from nonlinearity and the stabilizing action of the viscosity so that the local solutions can not be extended globally in time. For small initial data the decisive role is played by viscosity, while for larger initial data the nonlinear term is strong enough to destroy regularization of the solutions through the main part operator. We will try to see such effect through the estimates obtained in the paper. 

In 2-D, our idea is to {\it improving a bit} the viscosity term to make the whole problem {\it subcritical}, such that the improved viscosity together with the known $L^2(\Omega)$ a priori estimate will control the nonlinear term. The way for obtaining such effect is to replace the classical viscosity term $Au = -\nu P\Delta u$ through a bit higher fractional diffusion $-A^{1+\alpha} u$, with small $\alpha >0$. Next, we will study the process of letting $\alpha$ to $0^+$;  which  properties/estimates of the solutions of regularized problems (with $\alpha >0$) are lost in such a limit. We also look at 3-D N-S equation as a {\it super-critical problem} in a sense that a stronger diffusion term, with $-P\Delta$ operator in the power $s> \frac{5}{4}$, is needed to guarantee the control on the nonlinear term with the use of the standard $L^2(\Omega)$ a priori estimate.

In 2-D, instead of \eqref{NS-equation} we consider a {\it family of sub-critical problems}, with $\alpha \in (0, \frac{1}{2}]$: 
\begin{equation}\label{fractal}
\begin{split}
&u_t = -A^{1+\alpha}u - P(u \cdot \nabla)u + Pf,   \quad t>0,  \\
&u(0,x) = u_0(x).
\end{split}
\end{equation}
The approximation proposed in 3-D is given in \eqref{reg} with $s> \frac{5}{4}$.   

Until very recently, fractional power operators were not seriously studied in the literature, therefore we hope the regularizations proposed will help to understand the difficulties faced in the original N-S equation. An analogous phenomena was studied recently \cite{D-K-S1} for the {\it quasi-geostrophic equation in $R^2$} (e.g. \cite{C-C-W, C-C, C-C1, W1, W2}). The technique used in the present paper is similar to that of \cite{D-K-S1}.

There were several tries of replacing the classical N-S equation, or the viscosity term in it, with another equation having better properties of solutions, starting with J. Leray {\it $\alpha$-regularization} reported in paper \cite{L}, see also \cite{F-H-T}. Modification of one factor in nonlinearity, using mollifier, was sufficient to improve properties of solutions.  Another modification of the N-S equation was proposed by J.-L. Lions in \cite[Chapter 1, Remarque 6.11]{Li}, where the $-\Delta$ operator was replaced with $-\Delta + \kappa(-\Delta)^l$, $l \geq \frac{5}{4}$. Further modifications can be  easily found in the literature. In fact, we follow here the idea of J.-L. Lions to replace the diffusion term with a stronger one fractal diffusion term.

\subsection{Introductory facts.} 
\noindent   
{\it Notation.} We are using standard notation for Sobolev spaces. Compare  \cite{TR} or \cite[Chapter 1]{C-D} for properties of fractional order Sobolev spaces; see also \cite{HE1} for Sobolev type embeddings. For $r \in \mathbb{R}$, let $r^-$ denotes a number strictly less than $r$ but close to it. Similarly, $r^+ >r$ and $r^+$ close to $r$. When needed for clarity of the presentation, we mark the dependence of the solution $u$ of \eqref{fractal} on $\alpha \in (0, \frac{1}{2}]$, calling it $u^\alpha$, or $\ue$ for the approximation in 3-D. 

Recall first \cite[Chapter 3]{C-D}, that studying in a Banach space $X$ an abstract Cauchy's problem  with {\it sectorial positive operator} $A$ and solutions varying in the phase space $X^\beta = D(A^\beta)$:   
\begin{equation}\label{absteq}
\begin{split}
&u_t + Au = F(u),   \\
&u(0) = u_0,
\end{split}
\end{equation}
knowing an {\it a priori estimate} of all its  potential $X^\beta$ solutions; $\|u(t)\|_Y \leq const$ in another Banach space $D(A)\subset Y$, we say that the nonlinear term $F$ is {\it sub-critical relative to such a priori bound}, if for each such $X^\beta$ solution $u(t)$ an estimate is valid 
\begin{equation}
\exists_{\theta \in (0,1)} \exists_{nondecreasing g:[0,\infty) \to [0,\infty)} \quad  \|F(u(t)\|_X \leq g(\|u(t)\|_Y) \bigl(1 + \|u(t)\|^\theta_{X^\beta}\bigr),  
\end{equation}
for all $t \in (0,\tau_{u_0})$, where $\tau_{u_0}$ is the 'life time' of that solution. In that case (e.g. \cite[Chapter 3]{C-D}) the $X^\beta$ norm of the local solution will be bounded on $[0,\tau_{u_0})$, which allows to extend such solution globally in time. If the above estimate is possible for $\theta=1$, but not for $\theta <1$, the nonlinearity $F$ is called {\it critical relative to that a priori estimate}. 

Note that critical nonlinearities are 'of the same order' in the equation as the main part operator $A$ (compare   \cite{WA}). The main part operator $A$ will {\it not control} the nonlinearity in that case, unless we find a better a priori estimate.   

\subsection{Properties of the operator $-\Delta$.}
Familiar in the theory of the N-S equation are the following spaces:
\begin{equation}
\begin{split}
&\mathcal{L}^r(\Omega) = [L^r(\Omega)]^3,  \\
&\mathcal{W}^{2,r}(\Omega) = [W^{2,r}(\Omega)]^3,  \\ 
&X_r = cl_{\mathcal{L}^r(\Omega)}\{\phi \in [C^\infty_0(\Omega)]^3; div \phi=0\},
\end{split}
\end{equation}
$1<r<\infty$. We define also the Stokes operator 
\begin{displaymath}
A_r = -\nu P_r \left[
\begin{array}{ccc}
\Delta & 0 & 0  \\
0  & \Delta  & 0  \\
0  &  0  &  \Delta 
\end{array}
\right], 
\end{displaymath} 
where $P_r$ denotes the projection from $\mathcal{L}^r(\Omega)$ to $X_r$ given by the decomposition of $\mathcal{L}^r(\Omega)$ onto the space of divergence-free vector fields and scalar-function gradient (e.g. \cite{TE1}). It is further known \cite[Lemma 1.1]{G-M}, that 
\begin{proposition}
The operator $-A_r$ considered with the domain
$$
D(A_r) = X_r \cap \{\phi \in \mathcal{W}^{2,r}(\Omega); \phi=0 \ \text{on} \ \partial\Omega\},
$$
generates on $X_r$ an {\em analytic semigroup} $\{ e^{-t A_r}\}$ for arbitrary $1<r<\infty$.
\end{proposition}

A complete description of the domains of fractional powers of the Stokes operator $A_r =-\nu P_r\Delta$, $D(A_r) = D(-\Delta) \cap X_r$, can be found in \cite[p. 269]{G-M}, or in \cite{GI1}. Note further, that the domains of negative powers of the operator $A_r$ are introduced through the relation (e.g. \cite[p.269]{G-M}):
$$
D(A_r^\beta) = D(A_{r'}^{-\beta})^*,  \  \frac{1}{r} + \frac{1}{r'} =1.
$$
Thus, $D(A_r^\beta), \beta<0$, is the completion of $X_r$ under the norm $\|A_r^\beta \cdot\|_{0,r'}$.    

It is also easy to see, that the resolvent of the operator $A_2$ fulfills an estimate: $\Re(\sigma(A_2)) \geq \nu\lambda_1$, where $\lambda_1$ is the first positive eigenvalue of $-\Delta$ in $L^2(\Omega)$ considered with the Dirichlet boundary condition. The same estimate remains valid for the operator considered in $L^r(\Omega)$ with any $r\in (1,\infty)$. It follows further from \cite[Lem. 3.1]{G-M}, that the resolvent of $A_r$ is compact, also the embeddings $D(A_r^\beta) \subset D(A_r^\alpha)$ are compact when $0 <\alpha <\beta$ (\cite[Th. 1.4.8]{HE}). In fact the operator $A=A_2$ (we skip the subscript further for simplicity) is self-adjoint in the Hilbert space $\mathcal{L}^2(\Omega)$; see e.g. \cite{GI, G-M}. For such type operators the powers of the order $(1+\alpha)$ have similar properties; in particular they are also sectorial operators.    

Consequently, the operators $A_r, 1<   r< \infty$, are {\it sectorial positive}.{\it Fractional powers} of the order $1+\alpha$ for such operators are introduced through the Balakrishnan formula (\cite{KO}):
\begin{equation}\label{generalexpression}
A^{1+\alpha} \phi = \frac{2 \sin(\pi\alpha)}{\alpha\pi} \int_0^\infty \lambda^\alpha [A(\lambda+A)^{-1}]^2 \phi d\lambda,  \  \phi\in D(A^2), 
\end{equation}

We recall that in case of the N-S equation the, specific for that problem, {\it a priori estimate} is obtained multiplying \eqref{NS-equation} in $[L^2(\Omega)]^3$ through $u$, to obtain:
\begin{equation}
\frac{1}{2} \frac{d}{dt} \| u \|_{\mathcal{L}^2(\Omega)}^2 = -\nu \| \nabla u \|_{\mathcal{L}^2(\Omega)}^2 - \int_{\Omega}\nabla p \cdot u dx + \int_{\Omega} f \cdot u dx,
\end{equation}
since the nonlinear component vanish in that calculation due to condition $div u =0$:
\begin{equation}
\sum_{j=1}^3 \int_{\Omega} \sum_{i=1}^3 u_i \frac{\partial u_j}{\partial x_i} u_j dx = -\frac{1}{2} \int_{\Omega} \sum_{i=1}^3 \frac{\partial u_i}{\partial x_i} \sum_{j=1}^3 u_j^2 dx = 0.
\end{equation}
The term $\int_{\Omega} \nabla p \cdot u dx$, for regular solutions, is transformed as follows:
\begin{equation}\label{L2}
\begin{split}
\int_{\Omega} \nabla p \cdot u dx &= \int_{\Omega} \bigl(\frac{\partial p}{\partial x_1} u_1 + \frac{\partial p}{\partial x_2} u_2 + \frac{\partial p}{\partial x_3} u_3 \bigr) dx   \\ 
&= -\int_{\Omega} p \, div u dx =0.
\end{split}
\end{equation}
Consequently, an $\mathcal{L}^2(\Omega)$ estimate is obtained: 
\begin{equation}\label{L2est}
\frac{1}{2} \frac{d}{dt} \| u \|_{\mathcal{L}^2(\Omega)}^2 = -\nu \| \nabla u \|_{\mathcal{L}^2(\Omega)}^2 + \int_{\Omega} f \cdot u dx \leq -\frac{\nu}{2} \| \nabla u \|_{\mathcal{L}^2(\Omega)}^2 + c_{\nu} \| f \|_{\mathcal{L}^2(\Omega)}^2 \leq -\frac{\nu}{2 c_P} \| u \|_{\mathcal{L}^2(\Omega)}^2 + c_{\nu} \| f \|_{\mathcal{L}^2(\Omega)}^2, 
\end{equation}
thanks to the {\it Poincar\'{e} inequality}. We finally obtain a global in time estimate of the solution:  
\begin{equation}\label{1.9}
\| u(t) \|_{\mathcal{L}^2(\Omega)}^2 \leq \max\{ \| u_0 \|_{\mathcal{L}^2(\Omega)}^2; \frac{2c_\nu c_P \| f \|_{\mathcal{L}^2(\Omega)}^2}{\nu}\},  
\end{equation}
where $c_P$ denotes the constant in the Poincar\'{e} inequality. Having already the last estimate one can return to \eqref{L2est} to see that
\begin{equation}\label{H1es}
\| u \|^2_{L^2(0,T; [H^1_0(\Omega)]^3)} \leq \frac{1}{2\nu} \bigl(c_\nu T \| f \|_{\mathcal{L}^2(\Omega)}^2 + \| u_0 \|^2_{\mathcal{L}^2(\Omega)} \bigr),  
\end{equation}
for arbitrary $T>0$.  These are the strongest natural a priori estimates that can be obtained for, sufficiently regular, solutions of the N-S equation. 

\begin{remark}
Note that similar estimates are also valid for solutions of the fractal approximations \eqref{fractal}, with only one difference that the $[H^1_0(\Omega)]^3$ norm is replaced with the $[H^{1+\alpha}_0(\Omega)]^3$ norm. Consequently, the estimates \eqref{1.9} and \eqref{H1es} are valid for all the solutions $u^\alpha$ {\em uniformly in $\alpha >0$}. 
\end{remark}

\subsection{Local in time solvability of the 3-D and 2-D N-S problems.}\label{loc} 
We will rewrite N-S equation in a form of an abstract parabolic equation with sectorial positive operator and solve it using Dan Henry's approach (\cite{HE, C-D}). I want to make a comment here, that using Henry's approach we have fairly large choice of {\it phase space} (= space where solution varies). In fact, calling here an interesting paper \cite{BE}, as far as we consider the corresponding to \eqref{absteq} linear Cauchy problem with sectorial positive operator:  
\begin{equation*}
\begin{split}
&u_t + Au =0, \  t>0,  \\
&u(0)= u_0,
\end{split}
\end{equation*}
we can {\it 'set it'} at any level of the {\it fractional power scale} $X^\beta = D((-A)^\beta), \beta \in \R$, corresponding to $-A$ (see \cite[Section V.2]{AM} for an extension of that idea). When we move to the {\it semilinear problem} \eqref{absteq} with nonlinearity $F$ subordinated to $A$, it is an art to choose the proper level at that scale to be the phase space for semilinear problem. For that, we need to consider a priori estimates available for the specific equation, usually of physical origin, e.g. following from energy decay or conservation of mass valid in the process described through the equation. The full semilinear problem will be next written abstractly as
\begin{equation*}
\begin{split}
&u_t + Au = F(u), \  t>0,  \\
&u(0)= u_0,
\end{split}
\end{equation*}

The standard way to set the problem in the above setting, in $L^2(\Omega)$ (see e.g. \cite{TE1, GI, G-M}), is to apply to the equation the projector $P= P_2 : [L^2(\Omega)]^3 \to H$, where $H$ is the closure in $[L^2(\Omega)]^3$ of the set of divergence free functions $\{u\in [C^\infty_0(\Omega)]^3; div u =0\}$. The pressure term disappears then from the equation. The realization $A$ of the diagonal matrix operator $\nu P[-\Delta]_{3\times 3}$ acts from $D(A) \to H$. We also introduce the energy space $V = \{u \in [H^1_0(\Omega)]^3; div u = 0\}$, and the simplified notation for the nonlinearity: $F(u)= -P(u\cdot \nabla)u$. 

Operator $A=A_2$ has an associated scale of {\it fractional order spaces} $X^\beta \subset [H^{2\beta}(\Omega)]^3, \beta \geq 0$. The realizations of $A$ in $X^\beta$ act from $D(A^\beta) = X^{\beta +1} \to X^{\beta}$ and are {\it sectorial  positive} operators (see e.g. \cite{TR, GI, HE, M-S}). 

 We will rewrite the classical N-S equation in an equivalent form, using the property of the divergence-free functions. We have:
\begin{equation}\label{equivalent}
\begin{split}
&u_{jt} = \nu \Delta u_j - \frac{\partial p}{\partial x_j} - \sum_{i=1}^3 \frac{\partial (u_i u_j)}{\partial x_i} + f_j,   \quad div u = 0, \quad x \in \Omega,\, t>0, j=1,2,3, \\
&u = 0, \quad t>0, \quad x \in \partial \Omega,   \\ 
&u(0,x) = u_0(x),
\end{split}
\end{equation}
where $u = (u_1,u_2,u_3)$.  

We will recall next an estimate, important for the further calculations, borrowed from \cite[Lemma 2.1]{G-M}. A similar observation was given also in \cite[p.18]{HE} in dimension one. 
\begin{corollary}\label{Giga}
For each $j, 1\leq j\leq N$, the operator $A^{-\frac{1}{2}} P \frac{\partial }{\partial x_j}$ extends uniquely to a bounded linear operator from $[L^r(\Omega)]^N$ to $X_r, 1<r<\infty$.
Consequently, the following estimate holds:
\begin{equation}
\| A^{-\frac{1}{2}} P(u\cdot\nabla) v\|_{[L^r(\Omega)]^N} \leq M(r) \| |u| |v|\|_{[L^r(\Omega)]^N}. 
\end{equation}
\end{corollary}

\begin{observation}\label{gmme}
We have also the following estimate used further in the text. From \cite[Lemma 2.1]{G-M} we get, for all $N \in \mathbb{N}$:  
\begin{equation}\label{gme}
\begin{split}
&\| A^{-\frac{1}{2}} P(u\cdot \nabla)v\|_{[L^2(\Omega)]^N} \leq c \||u| |v|\|_{[L^2(\Omega)]^N} \leq c \|u\|_{[L^4(\Omega)]^N} \|v\|_{[L^4(\Omega)]^N},  \\
&\| P(u\cdot \nabla)v\|_{[L^2(\Omega)]^N} \leq c \| u \|_{[L^4(\Omega)]^N} \| \nabla v \|_{[L^4(\Omega)]^N}.
\end{split}
\end{equation}
Now, for any $\delta \in (0, \frac{1}{2})$, using the theory of interpolation;
\begin{equation}\label{2crit}
\begin{split}
\| A^{-\delta} P(u\cdot \nabla)v\|_{[L^2(\Omega)]^N} &\leq c \| A^{-\frac{1}{2}} P(u\cdot \nabla)v\|^{2\delta}_{[L^2(\Omega)]^N} \| P(u\cdot \nabla)v\|^{1-2\delta}_{[L^2(\Omega)]^N}   \\
&\leq c \| u \|^{1+ 2\delta}_{[L^4(\Omega)]^N} \| \nabla v \|^{1- 2\delta}_{[L^4(\Omega)]^N}.
\end{split}
\end{equation} 

In a similar way, starting from the estimates:
\begin{equation}
\begin{split}
&\| A^{-\frac{1}{2}} P(u\cdot \nabla)v\|_{[L^2(\Omega)]^N} \leq c \||u| |v|\|_{[L^2(\Omega)]^N} \leq c \|u\|_{[L^6(\Omega)]^N} \|v\|_{[L^3(\Omega)]^N},  \\
&\| P(u\cdot \nabla)v\|_{[L^2(\Omega)]^N} \leq c \| u \|_{[L^6(\Omega)]^N} \| \nabla v \|_{[L^3(\Omega)]^N},
\end{split}
\end{equation}
for any $\delta \in (0, \frac{1}{2})$, we get:
\begin{equation}\label{3crit}
\begin{split}
\| A^{-\delta} P(u\cdot \nabla)v\|_{[L^2(\Omega)]^N} &\leq c \| A^{-\frac{1}{2}} P(u\cdot \nabla)v\|^{2\delta}_{[L^2(\Omega)]^N} \| P(u\cdot \nabla)v\|^{1-2\delta}_{[L^2(\Omega)]^N}   \\
&\leq c \| u \|_{[L^6(\Omega)]^N} \| u \|^{2\delta}_{[L^3(\Omega)]^N} \| \nabla v \|^{1- 2\delta}_{[L^3(\Omega)]^N}.
\end{split}
\end{equation}
\end{observation}
The above estimates \eqref{2crit}, \eqref{3crit}, are valid for all the space dimensions $N= 2,3,4,...$. They can be extended further, using Sobolev type estimates, in a way depending on $N$.

For local in time solvability, we will set the problem \eqref{equivalent} in the base space $X^{-\frac{1}{4}}$ for the space dimension $N=2$, and in the base space $X^{-\frac{1}{8}}$ for the space dimension $N=3$. The corresponding {\it phase spaces} will be; $X^{\frac{1}{2}^+} \subset [H^{1^+}(\Omega)]^2)$ in case $N=2$, and $X^{\frac{3}{4}^+} \subset [H^{\frac{3}{2}^+}(\Omega)]^3)$ in case $N=3$ (e.g. \cite[Proposition 1.4]{G-M}). Note that, in both cases, the phase spaces are contained in the space $[L^\infty(\Omega)]^N$. Note also, there is another possible choice of the phase spaces (e.g. \cite{G-M}), if we decide to work in the spaces $[L^r(\Omega)]^N, N=2,3$, with $r>N$. 

We will formulate now the corresponding local existence results for $N =2,3$. 

{\bf Case $N=3$}. The main tool is the estimate taken from \cite[Lemma 2.2]{G-M} (with $\delta= \frac{1}{8}, \theta=\rho= \frac{3}{4}$):
\begin{equation}\label{1.14}
\| A^{-\frac{1}{8}} P(u\cdot\nabla)v\|_{[L^2(\Omega)]^3} \leq M \|A^{\frac{3}{4}^+} u\|_{[L^2(\Omega)]^3} \|A^{\frac{3}{4}^+} v\|_{[L^2(\Omega)]^3}.
\end{equation}
Since the form above is bi-linear, we have also the following consequences of the last estimate:
\begin{equation}
\begin{split}
&\| A^{-\frac{1}{8}} P((u-v)\cdot\nabla)v\|_{[L^2(\Omega)]^3} \leq M \|A^{\frac{3}{4}^+} (u-v)\|_{[L^2(\Omega)]^3} \|A^{\frac{3}{4}^+} v\|_{[L^2(\Omega)]^3},  \\
&\| A^{-\frac{1}{8}} P(u\cdot\nabla)(u-v)\|_{[L^2(\Omega)]^3} \leq M \|A^{\frac{3}{4}^+} u \|_{[L^2(\Omega)]^3} \|A^{\frac{3}{4}^+} (u-v)\|_{[L^2(\Omega)]^3}.
\end{split}
\end{equation}

Consequently, the nonlinear term $F(u) +Pf = -P(u\cdot \nabla)u + Pf$ acts from $D(A^{\frac{3}{4}^+}) \subset [H^{\frac{3}{2}^+}(\Omega)]^3$ into $D(A^{-\frac{1}{8}})$ as a map, Lipschitz continuous on bounded subsets of $D(A^{\frac{3}{4}^+})$. According to \cite{HE, C-D}, this suffices to obtain a {\it local in time solution} of the 3-D equation \eqref{NS-equation} (also of 3-D equation \eqref{reg}), more precisely:    
\begin{theorem}\label{localexistence} 
When $Pf \in D(A^{-\frac{1}{8}})$, $u_0 \in D(A_2^{\frac{3}{4}^+})$, then there exists a unique local in time {\em mild solution} $u(t)$ to \eqref{NS-equation} in the phase  space $D(A^{\frac{3}{4}^+}) \subset [H^{\frac{3}{2}^+}(\Omega)]^3$. Moreover, 
\begin{equation}
u \in C([0,\tau); D(A^{\frac{3}{4}^+})) \cap C((0,\tau); D(A^{\frac{7}{8}})) , \  u_t \in C((0,\tau); D(A^{\frac{7}{8}^-})).
\end{equation}
Here $\tau> 0$ is the 'life time' of that local in time solution. Moreover, the {\em Cauchy formula} is satisfied:
\begin{equation*}
u(t) = e^{-A t} u_0 + \int_0^t e^{-A(t-s)} F(u(s)) ds, \  t \in [0, \tau), 
\end{equation*}
where $e^{-A t}$ denotes the linear semigroup corresponding to the operator $A$. We need also to mention that the considered here mild solutions have additional regularity properties, as described in particular in \cite[p.218]{C-D}; here $u \in C^1((0,\tau); D(A^{\frac{7}{8}^-}))$. This property is used in the calculations below.      
\end{theorem}  

{\bf Case $N=2$}. We will use a version of the estimate in \cite[Lemma 2.2]{G-M} (with $\delta= \frac{1}{4}, \theta=\rho= \frac{1}{2}$):
\begin{equation}
\| A^{-\frac{1}{4}} P(u\cdot \nabla)v\|_{[L^2(\Omega)]^2} \leq M \|A^{\frac{1}{2}^+} u\|_{[L^2(\Omega)]^3} \|A^{\frac{1}{2}^+} v\|_{[L^2(\Omega)]^2}.
\end{equation}
Since the form above is bi-linear, we have also the following consequences of the last estimate:
\begin{equation}
\begin{split}
&\| A^{-\frac{1}{4}} P((u-v)\cdot\nabla)v\|_{[L^2(\Omega)]^2} \leq M \|A^{\frac{1}{2}^+} (u-v)\|_{[L^2(\Omega)]^2} \|A^{\frac{1}{2}^+} v\|_{[L^2(\Omega)]^2},  \\
&\| A^{-\frac{1}{4}} P(u\cdot\nabla)(u-v)\|_{[L^2(\Omega)]^2} \leq M \|A^{\frac{1}{2}^+} (u-v)\|_{[L^2(\Omega)]^2} \|A^{\frac{1}{2}^+} (u-v)\|_{[L^2(\Omega)]^2}.
\end{split}
\end{equation}
Consequently, the nonlinear term $F(u)+ Pf= -P(u\cdot\nabla)u + Pf$ acts from $D(A^{\frac{1}{2}^+}) \subset [H^{1^+}(\Omega)]^2$ into $D(A^{-\frac{1}{4}})$ as a map, Lipschitz continuous on bounded subsets of $D(A^{1^+})$. According to \cite{HE, C-D}, this suffices to obtain a {\it local in time solution} of the 2-D equation \eqref{NS-equation} (also of 2-D equation \eqref{fractal}), more precisely:    
\begin{theorem}\label{localexistence'} 
When $Pf \in D(A^{-\frac{1}{4}})$, $u_0 \in D(A^{\frac{1}{2}^+})\subset [H^{1^+}(\Omega)]^2$, then there exists a unique local in time {\em mild solution} $u(t)$ to \eqref{NS-equation} in the phase  space $D(A^{\frac{1}{2}^+}) \subset [H^{1^+}(\Omega)]^2$. Moreover, 
\begin{equation}\label{1.24}  
u \in C([0,\tau); D(A^{\frac{1}{2}^+})) \cap C((0,\tau); D(A^{\frac{3}{4}})) , \  u_t \in C((0,\tau); D(A^{\frac{3}{4}^-})).
\end{equation}
Here $\tau> 0$ is the 'life time' of that local in time solution. Moreover, the {\em Cauchy formula} is satisfied:
\begin{equation*}
u(t) = e^{-A t} u_0 + \int_0^t e^{-A(t-s)} F(u(s)) ds, \  t \in [0, \tau), 
\end{equation*}
where $e^{-A t}$ denotes the linear semigroup corresponding to the operator $A$.     
\end{theorem}

\subsection{Super-criticality of the N-S equation in 3-D, and criticality in 2-D}\label{supercrit} 
We will consider now {\it criticality} of the N-S equation \eqref{NS-equation} in the sense \cite{WA, C-D}, that means {\it 'the nonlinear term is of the same order as the main part operator'}.    

{\bf Case $N=3$}. The 3-D N-S equation is {\it super-critical} in the sense stated below. A general problem \eqref{absteq} will be called {\it super-critical with respect to the a priori estimate in $Y$}, if for all its possible $X^\beta$ solutions an estimate holds
\begin{equation}\label{supercritical}
\exists_{\theta_0 >1} \forall_{\theta > \theta_0} \exists_{nondecreasing g:[0,\infty) \to [0,\infty)}  \quad  \|F(u(t)\|_X \leq g(\|u(t)\|_Y) (1 + \|A^\beta u(t)\|^\theta_X),  
\end{equation}
for all $t \in (0,\tau_{u_0})$, and such type estimate {\it is not true} for exponents $\theta< \theta_0$.   

In case of the Navier-Stokes equation \eqref{NS-equation} the mentioned above a priori estimate will be the $Y=[L^2(\Omega)]^N$ estimate \eqref{1.9}.    

We will see that, as a consequence of \eqref{3crit}, {\it the problem \eqref{NS-equation} is  super-critical}. Indeed, the estimate \eqref{3crit} written for the local solution $u=u(t)$ obtained in Theorem \ref{localexistence} extends for $N=3$, with the use of the Nirenberg-Gagliardo type estimates, to:  
\begin{equation}\label{1.26}
\| A^{-\frac{1}{8}} P(u\cdot\nabla)u\|_{[L^2(\Omega)]^3} \leq c \| u \|_{[L^6(\Omega)]^3} \| u \|^{\frac{1}{4}}_{[L^3(\Omega)]^3} \| \nabla u \|^{\frac{3}{4}}_{[L^3(\Omega)]^3} 
\leq c \| u \|^{\frac{9}{7}}_{[H^{\frac{7}{4}}(\Omega)]^3} \| u \|^{\frac{5}{7}}_{[L^2(\Omega)]^3}.
\end{equation}
This shows the nonlinearity of the 3-D N-S equation is super-critical as a map from $X^{\frac{7}{8}}$ to $X^{-\frac{1}{8}}$. Exponent $\theta$ obtained above equals $\frac{9}{7}$. 

Using the estimate of Corollary \ref{Giga} we will find now the value $\theta_0$ in case of the 3-D N-S equation. We obtain,
\begin{equation}
\begin{split}
\| P(u\cdot \nabla)u\|_{X^{-\frac{1}{2}}} & = \| A^{-\frac{1}{2}} P(u\cdot \nabla)u\|_{[L^2(\Omega)]^3} \leq M \| |u|^2\|_{[L^2(\Omega)]^3} \leq c \| u \|^2_{[L^4(\Omega)]^3} \\
& \leq c \| u \|^{\frac{3}{2}}_{[H^1(\Omega)]^3} \| u\|^{\frac{1}{2}}_{[L^2(\Omega)]^3},  
\end{split}
\end{equation}
and the estimates are sharp.  Therefore, $\theta_0 =\frac{3}{2}$ in that case. 

Finally, for the further use, we will check how large the exponent $s>1$ should be, for the 'strengthen diffusion' of the form $(-\Delta)^s$ replacing the usual $(-\Delta)$ operator in \eqref{NS-equation} to make the $L^2(\Omega)$ estimate critical. Using again Corollary \ref{Giga} we get that 
\begin{equation}\label{criticals}
\| A^{-\frac{1}{2}} P(u\cdot \nabla)u\|_{[L^2(\Omega)]^3} \leq M \| |u|^2\|_{[L^2(\Omega)]^3} \leq c \| u \|^{\frac{3}{2(2s-1)}}_{[H^{2s-1}(\Omega)]^3} \| u \|^{\frac{4s-5}{4s-2}}_{[L^2(\Omega)]^3}.
\end{equation}
We find such critical value of $s$ from the condition; $\frac{3}{2(2s-1)}=1$. Consequently, $s = \frac{5}{4}$.

{\bf Case $N=2$}. Using Corollary \ref{Giga} it is easy to see {\it criticality of the 2-D Navier-Stokes equation}.  More precisely, to verify that its nonlinearity is critical with respect to the standard $L^2(\Omega)$ a priori estimates, as a map from $X^{\frac{1}{2}} \subset [H^1(\Omega)]^2$ to $X^{-\frac{1}{2}}$. Indeed, 
\begin{equation}
\begin{split}
\| P(u\cdot \nabla)u\|_{X^{-\frac{1}{2}}} &\leq \| A^{-\frac{1}{2}} P(u\cdot \nabla)u\|_{[L^2(\Omega)]^2} \leq M \| |u|^2\|_{[L^2(\Omega)]^2}    \\
&\leq c \| u \|^2_{[L^4(\Omega)]^2} \leq \| u \|_{[H^1(\Omega)]^2} \| u \|_{[L^2(\Omega)]^2} = c(\|u\|_{[L^2(\Omega)]^2}) \| u \|^1_{[H^1(\Omega)]^2},
\end{split}
\end{equation} 
and {\it no better estimate (with exponent smaller than 1) is possible}. 

\begin{observation}
Using the estimate \eqref{2crit} with $\delta = \frac{1}{4}$, we will show that the nonlinearity in 2-D N-S equation is {\em critical as a map from $X^{\frac{3}{4}} \subset [H^{\frac{3}{2}}(\Omega)]^2$ to $X^{-\frac{1}{4}}$}. Indeed, from \eqref{2crit} with $\delta = \frac{1}{4}, N =2$,
\begin{equation}
\| A^{-\frac{1}{4}} P(u\cdot \nabla)u\|_{[L^2(\Omega)]^2} \leq c \| u \|^{\frac{3}{2}}_{[L^4(\Omega)]^2} \| \nabla u \|^{\frac{1}{2}}_{[L^4(\Omega)]^2} \leq c \| u \|_{[L^2(\Omega)]^2} \| u \|_{[H^{\frac{3}{2}}(\Omega)]^2}, 
\end{equation}
 where the Nirenberg-Gagliardo type estimates were used:
 \begin{equation}
 \begin{split}
 &\| \phi \|_{L^4(\Omega)} \leq c \| \phi \|^{\frac{1}{3}}_{H^{\frac{3}{2}}(\Omega)} \| \phi \|^{\frac{2}{3}}_{L^2(\Omega)},  \\
 &\| \phi \|_{W^{1,4}(\Omega)} \leq c \| \phi \|_{H^{\frac{3}{2}}(\Omega)}.
\end{split}
\end{equation}
No better estimates are possible. 
\end{observation}

Note further, that in 2-D the nonlinearity is also critical with respect to the $L^2(\Omega)$ a priori estimates as a map between $X^1= D(A) \subset [H^2(\Omega)]^2$ and $X_2$. Indeed, the following estimate holds,
\begin{equation}
\| P(u\cdot \nabla)u\|_{X_2} \leq c \| u\|_{[L^4(\Omega)]^2} \|\nabla u\|_{[L^4(\Omega)]^2} \leq c \| u \|_{[H^2(\Omega)]^2} \| u \|_{[L^2(\Omega)]^2},
\end{equation}
and {\it no smaller exponent on the $H^2(\Omega)$ norm is possible}.

\section{Global in time solutions in 3-D. Small data.} 
As well known, global in time extendibility of the local mild solution constructed in the Theorem \ref{localexistence} is possible provided we have sufficiently well a priori estimates that prevents the $D(A^{\frac{3}{4}^+})\subset [H^{\frac{3}{2}^+}(\Omega)]^3$ norm of the solution to blow up in a finite time. We will propose next such type estimate, in $[H^1(\Omega)]^3$, for solutions of the 3-D N-S equation when the data; $u_0$ and $f$ are smooth and sufficiently small. Such estimate will be used later to construct global in time solutions of the N-S with small data. Another approach to that problem in arbitrary dimension $N$, using estimates on integral equation, was presented in \cite{C-D1}.    
\begin{theorem}
If $u_0 \in D(A) \subset[H^2(\Omega)]^3$ and $f \in [L^2(\Omega)]^3$ fulfill the 'smallness restriction' \eqref{smallrest}, then the $[H^1(\Omega)]^3$ norms of the solutions $u$ are bounded uniformly in time $t\geq 0$.  
\end{theorem}

\begin{proof}
It is known (e.g. \cite{HE, C-D}), that the local in time solutions $u$ are regularized for $t>0$ (through the equation) into the space $D(A)\subset [H^2(\Omega)]^3$. When the initial data $u_0 \in D(A)$, they simply vary in $D(A)$ for $t \geq 0$ small, until possible blow-up time $t(\alpha, u_0)$. We want to show that, to small $u_0$ and $f$ correspond small solutions, in the $[H^1(\Omega)]^3$ norm (uniformly in time).

To get  estimate of the solution $u$ in $[H^1(\Omega)]^3$ multiply \eqref{fractal} by $A u$ to obtain:
\begin{equation}
<u_t, Au>_{[L^2(\Omega)]^3} = -<Au, Au>_{[L^2(\Omega)]^3} - <P(u \cdot \nabla)u, Au>_{[L^2(\Omega)]^3} + <Pf, Au>_{[L^2(\Omega)]^3},
\end{equation}
since the pressure term vanishes. Thanks to \eqref{gme}, this gives,     
\begin{equation}
\begin{split}
\frac{1}{2} \frac{d}{dt} \| u \|^2_{[H^1(\Omega)]^3} &\leq -c\nu \| u \|^2_{[H^2(\Omega)]^3} + c\| u \|_{[W^{1,4}(\Omega)]^3} \| u \|_{[L^4(\Omega)]^3} \| u \|_{[H^2(\Omega)]^3} + \| Pf \|_{[L^2(\Omega)]^3} \| u \|_{[H^2(\Omega)]^3}  \\
&\leq -c\nu \| u \|^2_{[H^2(\Omega)]^3} + c_1 \| u \|^{\frac{7}{4}}_{[H^1(\Omega)]^3} \| u \|^{\frac{5}{4}}_{[H^2(\Omega)]^3} + \| Pf \|_{[L^2(\Omega)]^3} \| u \|_{[H^2(\Omega)]^3}  \\
&\leq -\frac{c' \nu}{2} \| u \|^2_{[H^1(\Omega)]^3} + C_\nu \bigl(\| u \|^{\frac{14}{3}}_{[H^1(\Omega)]^3} + \| Pf \|^2_{[L^2(\Omega)]^3}\bigr),
\end{split}
\end{equation}
with the standard use of Young's inequality, and the embedding $H^2(\Omega) \subset H^1(\Omega)$ (constant $c'$). Denoting; $y(t) := \| u(t) \|^2_{[H^1(\Omega)]^3}$, we arrive at the  {\it differential inequality} (e.g. \cite{Sz, Wa}):      
\begin{equation}\label{1.32}
\begin{split}
&\frac{1}{2} y'(t) \leq -\frac{c'\nu}{2} y(t) + C_\nu y^{\frac{7}{3}}(t) + C_\nu\| Pf \|^2_{[L^2(\Omega)]^3},  \\
&y(0) = \| u(0) \|^2_{[H^1(\Omega)]^3}.
\end{split}
\end{equation}
Analyzing its right hand side, real function $g(z) = -\frac{c'\nu}{2} z + C_\nu z^{\frac{7}{3}} + C_\nu \| Pf \|^2_{[L^2(\Omega)]^3}$, we see that $g(0)= C_\nu \| Pf \|^2_{[L^2(\Omega)]^3} >0, g'(0)<0$, and $g$ has a minimum for the argument $z_{min}= \bigl(\frac{3 c' \nu}{14 C_\nu}\bigr)^{\frac{3}{4}}$, with $g(z_{min})<0$ when the 'free term' $C_\nu\| Pf \|^2_{[L^2(\Omega)]^3}$ is small. More precisely, to keep the value of $y(t)$ bounded for all positive times, we need to assume the {\it smallness hypothesis}: {\it Let the data: $\| u(0) \|^2_{[H^1(\Omega)]^3}$ and $\| Pf \|^2_{[L^2(\Omega)]^3}$ be so small, that:
\begin{equation}\label{smallrest}
\begin{split}
&g(z_{min})< 0, \  \text{equivalently}  \  \| Pf \|^2_{[L^2(\Omega)]^3} < \frac{2c' \nu}{7 C_\nu}, \  \text{and}   \\   
&\| u_0\|^2_{[H^1(\Omega)]^3} \leq z_{min} = \left(\frac{3c' \nu}{14 C_\nu}\right)^{\frac{3}{4}}.   
\end{split}
\end{equation}
Note that $\bigl(\frac{\nu}{C_\nu}\bigr)^{\frac{3}{4}}$ is proportional to $\nu^2$. } Consequently we obtain the bound
\begin{equation}\label{bound}
\| u(t)\|^2_{[H^1(\Omega)]^3} \leq z_{min} \  \text{valid for all} \  t\geq 0. 
\end{equation}

With the last assumption, the smooth local solutions $u(t)$, introduced in Theorem \ref{globalsolutions} are bounded in $[H^1(\Omega)]^3$ uniformly in $t\geq 0$. Note, that a bit more accurate bounds in \eqref{smallrest} are possible if one compare the data with the two positive zeros of the function $g$. See also the corresponding restrictions formulated in \cite[Theorem 3.7]{TE1}.  
\end{proof} 

\subsection{Regularization of the 3-D N-S equation} In the 3-D case the N-S equation \eqref{NS-equation} is super-critical. Global in time extendibility of the local solutions constructed in Theorem \ref{localexistence} is in general not possible (unless for small data). The viscosity term in the equation \eqref{NS-equation}  is not strong enough to control the nonlinearity and the regularization used in 2-D case is also not sufficient to work in the 3-D case. In the present subsection we propose another approximation/regularization (with stronger diffusion term) of the 3-D N-S equation, for which the $L^2(\Omega)$ estimates (same as for the original equation) are sufficient to make such problems {\it subcritical}. 

We propose now approximation of the original 3-D Navier-Stokes equation having global in time, unique and regular solutions. Consider namely the approximation/regularization of \eqref{NS-equation} of the form:
\begin{equation}\label{reg}
\begin{split}
&u_t = -(A + \frac{\epsilon}{\nu^{s-1}}A^s)u - P(u \cdot\nabla)u + Pf,   \quad t>0,  \\
&u(0,x) = u_0(x),
\end{split}
\end{equation}
with a parameter $s>1$ (to be chosen), and $\epsilon>0$. For fixed (for a moment) parameter $\epsilon$, denote the solution to the above problem as $\eu$. We mean here the solution on the base space $D(A^{-\frac{1}{4}})$, obtained in a similar way as for the original N-S equation \eqref{NS-equation} in subsection \ref{loc}.   

It is clear how to determine the proper, sufficiently large, value of the exponent $s>1$  to guarantee, together with the $L^2(\Omega)$ estimates valid also for solutions of \eqref{reg}, that the nonlinear term is subordinated to the main part operator $(A + \frac{\epsilon}{\nu^{s-1}}A^s)$. We need to compare the bound obtained for the nonlinearity with the income from the improved viscosity term. Estimate of the nonlinearity, obtained from \eqref{2crit}, reads  
\begin{equation}\label{1.36}
\| A^{-\frac{1}{4}} F(\eu(t)) \|_{[L^2(\Omega)]^3} \leq c \|\eu(t)\|^{\frac{3}{2}}_{[L^4(\Omega)]^3} \|\eu(t)\|^{\frac{1}{2}}_{[W^{1,4}(\Omega)]^3}. 
\end{equation} 
Let $s> \frac{5}{4}$. With the use of the Nirenberg-Gagliardo type estimate we obtain the {\it subordination condition} 
\begin{equation}
\| A^{-\frac{1}{4}} F(\eu(t)) \|_{[L^2(\Omega)]^3} \leq c(\| \eu(t) \|_{[L^2(\Omega)]^3}) \| \eu(t)\|^\theta_{[H^{2s-\frac{1}{2}}(\Omega)]^3} \leq c'(\| \eu(t) \|_{[L^2(\Omega)]^3}) \| A^{s-\frac{1}{4}}\eu(t)\|^\theta_{[L^2(\Omega)]^3}, 
\end{equation}
where $\theta= \frac{2}{2s-\frac{1}{2}} < 1$. Consequently, the standard $L^2(\Omega)$ a priori estimate is sufficient to assure the global in time extendibility of such local solutions $\eu$ when $s >\frac{5}{4}$. We will study such approximation next.  

\begin{remark}
The exponent $\frac{5}{4}$ proposed for regularization of \eqref{NS-equation} in \cite[Chapter 1, Remarque 6.11]{Li} is the same (also, as in \eqref{criticals}).   
\end{remark} 

To make further considerations simpler, let us consider the introduced above regularization of the original 3-D N-S equation {\it with exponent $s=2$ in \eqref{reg}} (which is a particular choice). We will describe next shortly the process of passing to the limit, as $\epsilon \to 0^+$, in such approximations. The idea is similar as in \cite{Li, DL, D-S}, using the {\it parabolic regularization technique}. An abstract counterpart of such regularization ($s=2$) of the 3-D N-S equation has the form:
\begin{equation}\label{sreg}
u_t = -Au - \frac{\epsilon}{\nu} A^2 u - P(u \cdot \nabla)u + Pf,  \quad  u(0) = u_0.
\end{equation}
Solutions of that problem will be denoted further as $\ue$. 

We formulate next the corresponding local existence result for such problem.  
\begin{theorem} \label{regth}
Consider \eqref{reg} with $s=2$ as equation in $D(A^{-\frac{1}{4}})$. When $Pf \in D(A^{-\frac{1}{4}})$, $u_0 \in D(A^{\frac{3}{2}}) \subset [H^3(\Omega)]^3$, then there exists a unique local in time {\em mild solution} $\ue(t)$ to \eqref{reg}, $s=2$, in the phase  space $D(A^{\frac{3}{2}})$. Moreover, 
\begin{equation}\label{2.10}
\ue \in C([0,\tau); D(A^{\frac{3}{2}})) \cap C((0,\tau); D(A^{\frac{7}{4}})) , \  \ue_t \in C((0,\tau); D(A^{\frac{7}{4}^-})).
\end{equation}
Here $\tau> 0$ is the 'life time' of that local in time solution. Also, the corresponding {\em Cauchy formula} is satisfied. 

Moreover, together with the standard $[L^2(\Omega)]^3$ a priori estimate valid for solutions of \eqref{reg} {\em uniformly in $\epsilon >0$}, the obtained above local solutions will be extended globally in time in the class \eqref{2.10}.  
\end{theorem}

The proof, similar as presented previously, is omitted. We only resolve the $L^2$ estimate for \eqref{sreg}, with an extra term compare to \eqref{L2est}. Multiplying \eqref{sreg} by $\ue$ we obtain
\begin{equation}
\begin{split}
<\ue_t, \ue>_{[L^2(\Omega)]^3} = -<A\ue, \ue>_{[L^2(\Omega)]^3} - &\frac{\epsilon}{\nu} <A^2 \ue, \ue>_{[L^2(\Omega)]^3}  \\ 
&- <P(\ue \cdot \nabla)\ue, \ue>_{[L^2(\Omega)]^3} + <Pf, \ue>_{[L^2(\Omega)]^3}.
\end{split}
\end{equation}
Noting that the nonlinear term vanishes, we get an estimate 
\begin{equation}
\frac{1}{2} \frac{d}{dt} \| \ue(t)\|^2_{[L^2(\Omega)]^3} \leq -\| A^{\frac{1}{2}}\ue\|^2_{[L^2(\Omega)]^3} - \frac{\epsilon}{\nu} \| A\ue \|^2_{[L^2(\Omega)]^3} + \| Pf \|_{[L^2(\Omega)]^3} \| \ue\|_{[L^2(\Omega)]^3}.
\end{equation}
The reasoning goes as for \eqref{L2est}, if we neglect the second right hand side term, giving precisely \eqref{1.9} for $\ue$. In addition to \eqref{H1es}, for arbitrary fixed $T>0$ we obtain an extra estimate 
\begin{equation}\label{H2e}
\epsilon \| \ue \|^2_{L^2(0,T; [H^2(\Omega)]^3)} \leq \frac{1}{2} \bigl(c_\nu T \| f \|_{\mathcal{L}^2(\Omega)}^2 + \| u_0 \|^2_{\mathcal{L}^2(\Omega)} \bigr).  
\end{equation}
The latter estimate will be used to pass to the limit, when $\epsilon \to 0^+$, in the weak formulation of \eqref{sreg}. The idea is taken from \cite[Chapter 3]{Li}.  

The problem \eqref{sreg} has a weak formulation; {\it for arbitrary  test function $v \in D(A)$ and fixed arbitrary $T>0$, we set}
\begin{equation}\label{weakk}
\begin{split}
\frac{d}{dt} <u, v>_{[L^2(\Omega)]^3} = -<A^{\frac{1}{2}}u, A^{\frac{1}{2}}v>_{[L^2(\Omega)]^3} &- \frac{\epsilon}{\nu} <Au, Av>_{[L^2(\Omega)]^3}   \\
&- <P(u \cdot \nabla)u, v>_{[L^2(\Omega)]^3} + <Pf, v>_{[L^2(\Omega)]^3}, 
\end{split}
\end{equation}
and the {\it weak solutions} are expected in $L^2(0,T; D(A))$. The solutions $\ue$ constructed in Theorem \ref{regth} fulfill the above weak form of \eqref{sreg}. The time derivative here will be understand in the sense of the 'scalar distributions' (e.g. \cite{Li, TE1}). Passing to the limit in the nonlinear term is possible thanks to the uniform estimates of $\ue$ in $L^2(0,T; [H^1_0(\Omega)]^3)$, as described below in 2-D case. Note also that, when letting $\epsilon \to 0^+$, the second right hand side term will tend to zero in the space $L^2(0,T; [H^1_0(\Omega)]^3)$, thanks to \eqref{H2e}. An even more interesting observation is that, when passed to such a limit, we obtain a 'weak solutions' of the original N-S equation (e.g. \cite[Chapter 3.1]{TE1}), due to J. Leray. The only difference is that the set of test functions $v\in D(A)$ we are using is smaller. But it is dense in $D(A^{\frac{1}{2}})$, so the two formulations are equivalent. We will not extend that considerations here in details.

\section{2-D critical N-S equation as a limit of sub-critical approximations.} We will describe now the convergence of the solutions of the fractal approximations \eqref{fractal} to the solution of the limiting 2-D N-S equation. Precisely as in Theorem \ref{localexistence'}, the solutions $\au$ of \eqref{fractal} (the superscript is added for clarity) will be constructed  in the class:  
\begin{equation}  
\au \in C([0,\tau); D(A^{\frac{1}{2}^+})) \cap C((0,\tau); D(A^{\frac{3}{4}})) , \  \au_t \in C((0,\tau); D(A^{\frac{3}{4}^-})).
\end{equation}
Further, the $L^2(\Omega)$ a priori estimates are satisfied for $\au$ uniformly in $\alpha \in (0, \frac{1}{2}]$. Consequently, we claim:  
\begin{theorem}\label{globalsolutions}
The local in time solution $\au(t)$ of the approximating problem  \eqref{fractal} constructed as in Theorem \ref{localexistence'} will be extended globally in time in the above class. Moreover, the standard $L^2(\Omega)$ a priori estimate \eqref{1.9} is valid for it {\em uniformly in $\alpha \in (0, \frac{1}{2}]$}.  
\end{theorem}
All the nice properties of the abstract semilinear sectorial equation, like elegant theory of existence and uniqueness, regularization of the solution for positive times, are valid for solutions $\au(t)$ with any $\alpha>0$. Typical regularization goes from the phase space $X^\beta, \beta<1$, to the space $X^1=D(A)$; see e.g. \cite{HE, C-D}. However, for problems \eqref{fractal}, estimates of norms better that $L^2(\Omega)$ will depend on $\alpha>0$, possibly blowing up when $\alpha \to 0^+$, and therefore {\it can not be extended in general} to a limit solution $u$.

\subsection{2-D \eqref{NS-equation} as a limit of \eqref{fractal} when $\alpha \to 0^+$.}\label{critical}    
A precise description of letting $\alpha \to 0^+$ in the equation \eqref{fractal} is given next. 
In this section we consider the solutions $\au$ of \eqref{fractal} constructed in Theorem \ref{globalsolutions} on the phase space $D(A^{\frac{3}{4}}) \subset [H^{\frac{3}{2}}(\Omega)]^2$. Such solutions, for any $\alpha \in (0,\frac{1}{2}]$, are varying in $[H^{\frac{3}{2}}(\Omega)]^2$. Indeed, according to \cite[Proposition 1.4]{G-M}: {\it For any $\beta \geq 0$, the domain $D(A^\beta)$ is continuously embedded in $X_2 \cap [H^{2\beta}(\Omega)]^2$}.  Solutions $\au$ fulfill also, {\it uniformly in $\alpha \in (0, \frac{1}{2}]$}, estimate \eqref{1.9} in $[L^2(\Omega)]^2$. More precisely, for such solutions of \eqref{fractal} we have an estimate:  
\begin{equation}
\exists_{const >0} \forall_{\alpha \in (0, \frac{1}{2}]}   \  \|\au \|_{L^\infty([0,\infty); [L^2(\Omega)]^2)} \leq const.  
\end{equation}
This is the {\it main information} allowing us to let $\alpha \to 0^+$ in the equation \eqref{fractal}. 

\bigskip \noindent  
{\bf Passing in \eqref{fractal} to the limit $\alpha \to 0^+$.} We look at \eqref{fractal} as an equation in $[L^2(\Omega)]^2$, and 'multiply' by the test function $A^{-1-\alpha}\phi$ where $\phi \in D(A^{\frac{3}{4}})$; recall that $D(A^{\frac{3}{4}})\subset [H^{\frac{3}{2}}(\Omega)]^2$,   
\begin{equation}\label{2.2}
<\au_t +P(\au \cdot\nabla)\au, A^{-1-\alpha}\phi>_{[L^2(\Omega)]^2}  = -<A^{1+\alpha}\au, A^{-1-\alpha}\phi>_{[L^2(\Omega)]^2} + <Pf, A^{-1-\alpha}\phi>_{[L^2(\Omega)]^2}. 
\end{equation}
We will discuss now the convergence of the terms in \eqref{2.2} one by one. Note that when $\alpha \to 0^+$ then, by Lemma \ref{new}, $A^{-1-\alpha}\phi \to A^{-1} \phi$.  Thanks to uniform in $\alpha \in (0,\frac{1}{2}]$ boundedness of $\au$ in $L^\infty([0,\infty);[L^2(\Omega)]^2)$, we obtain: 
\begin{equation}\label{mic}
\begin{split}
&<A^{1+\alpha}\au, A^{-1-\alpha}\phi>_{[L^2(\Omega)]^2} = <\au, \phi>_{[L^2(\Omega)]^2} \to <u, \phi>_{[L^2(\Omega)]^2},  \\
&<Pf, A^{-1-\alpha}\phi>_{[L^2(\Omega)]^2} \to <Pf, A^{-1} \phi>_{[L^2(\Omega)]^2},
\end{split}
\end{equation}
where $u$ is the weak limit of $\au$ in $[L^2(\Omega)]^2$ as $\alpha\to 0^+$ (over a sequence $\{\alpha_n\}$ convergent to $0^+$; various sequences may lead to various weak limits). 

We return to \eqref{2.2} to see that letting $\alpha \to 0^+$ over a sequence $\{\alpha_n\}$, where $u$ denotes weak limit in $[L^2(\Omega)]^2$ of such sequence, we have  
\begin{equation}
\lim_{\alpha_n \to 0} <A^{-\alpha}\bigl(\au_t + P(\au \cdot\nabla)\au\bigr), A^{-1}\phi>_{[L^2(\Omega)]^2} = -<u, \phi>_{[L^2(\Omega)]^2} + <Pf, A^{-1} \phi>_{[L^2(\Omega)]^2}, 
\end{equation}
since the right hand side is convergent. Consequently, the left hand side has a limit as $\alpha_n \to 0^+$,  
\begin{equation}\label{added}
\lim_{\alpha_n \to 0} <A^{-\alpha}\bigl(\au_t + P(\au \cdot\nabla)\au\bigr), A^{-1}\phi>_{[L^2(\Omega)]^2} = \omega_\phi. 
\end{equation}

Note that, for the 'test functions'  $A^{-1} \phi$ varying in a {\it separable} Banach space, passing countable many times to a subsequence, we can chose a common subsequence proper for all test functions in dense subset of the space. Consequently, the equation below will be fulfilled in the whole space. We obtain 
\begin{equation}
\forall_{\phi \in X_2}  \quad  \omega_\phi = -<u, \phi>_{[L^2(\Omega)]^2} + <Pf, A^{-1} \phi>_{[L^2(\Omega)]^2},
\end{equation} 
which is a weak form of the limiting equation.

{\bf Separation of terms.} The two terms of $[A^{-\alpha}\bigl(\au_t + P(\au \cdot\nabla)\au\bigr)]$ will be separated when letting $\alpha \to 0^+$. More precisely we have:     
\begin{remark}\label{explanation}
Since the approximating solutions $\au$ satisfy (in particular)  
\begin{equation}
\au \in L^\infty(0,T; [L^2(\Omega)]^2), \au_t \in L^2(0,T; [L^2(\Omega)]^2),  
\end{equation}
then by \cite[Lemma 1.1, Chapt.III]{TE1} 
\begin{equation}
\forall_{\eta \in X_2} <\au_t, \eta>_{[L^2(\Omega)]^2} = \frac{d}{dt}<\au, \eta>_{[L^2(\Omega)]^2} \to \frac{d}{dt}<u,\eta>_{[L^2(\Omega)]^2},
\end{equation}
the time derivative $\frac{d}{dt}$ and the convergence are understood in $\mathcal{D}'(0,T; [L^2(\Omega)]^2)$ (space of the 'scalar distributions' \cite{Li}). Consequently,
\begin{equation}
\omega_\phi = \frac{d}{dt} <A^{-\alpha}u, A^{-1}\phi>_{[L^2(\Omega)]^2} + \omega^1_\phi,
\end{equation}
where $\omega^1_\phi$ is a limit in $\mathcal{D}'(0,T; [L^2(\Omega)]^2)$ of $<P(\au \cdot\nabla)\au, A^{-1-\alpha}\phi>_{[L^2(\Omega)]^2}$ over a chosen sequence $\alpha_n \to 0^+$. 

Convergence of the nonlinear term $F(\au)=P(\au \cdot\nabla)\au$ will be discussed next. As seen from \eqref{H1es}, the approximating solutions $\au$ are bounded in $L^2(0,T; [H^1_0(\Omega)]^2)$ uniformly in $\alpha >0$. Estimate \eqref{gme}: 
\begin{equation}
\|A^{-\frac{1}{2}}F(\au)\|_{[L^2(\Omega)]^2} \leq c\| \au\|^2_{[L^4(\Omega)]^2} \leq c'\| \au \|_{[L^2(\Omega)]^2} \| \au \|_{[H^1_0(\Omega)]^2}, N=2,
\end{equation}
together with the consequence of equation \eqref{fractal} 
$$
A^{-(\frac{1}{2}+\alpha)} \au_t = -A^{\frac{1}{2}}\au + A^{-(\frac{1}{2}+\alpha)}F(\au) + A^{-(\frac{1}{2}+\alpha)} Pf \in L^2(0,T; [L^2(\Omega)]^2),
$$
show that $\au_t$ are bounded in $L^2(0,T; D(A^{-1}))$ uniformly in $\alpha \in (0,\frac{1}{2}]$. By {\em Lions compactness lemma} \cite[Theorem 5.1]{Li}, the family  $\{\au\}_{\alpha \in (0,\frac{1}{2}]}$ as bounded in the space
$$
W = \{ \phi; \phi \in L^2(0,T; [H^1_0(\Omega)]^2), \phi_t \in L^2(0,T; D(A^{-1})) \}, 
$$ 
is precompact in $L^2(0,T; [H^{1^-}_0(\Omega)]^2)$. In particular any sequence $\{u^{\alpha_n}\}, \alpha_n \to 0^+$, has a subsequence convergent almost everywhere in $(0,T)\times \Omega$. Moreover, the above compactness allows to pass to a limit (using estimates of Observation \ref{gmme}) in the nonlinear term in \eqref{2.2}. Indeed, 
if $u^{\alpha_n} \to u, \alpha_n \to 0^+$ in the above sense then, using a consequence of \cite[Lemma 2.2]{G-M}; for small $\epsilon>0$ 
\begin{equation*}
\|A^{-(\frac{1}{2}+\epsilon)} P[(u\cdot \nabla)u - (u^{\alpha_n}\cdot \nabla)u^{\alpha_n}] \|_{[L^2(\Omega)]^2} \leq C\| |u- u^{\alpha_n}|(|u| + |u^{\alpha_n}|)\|_{[L^s(\Omega)]^2},
\end{equation*}
where $\frac{1}{s}=\frac{1}{2} +\epsilon$ (so that $s<2$), we have 
\begin{equation}
\begin{split}
\int_0^T <P(u^{\alpha_n}\cdot\nabla) u^{\alpha_n}, A^{-1-\alpha_n} \phi>_{[L^2(\Omega)]^2} dt &= \int_0^T <A^{-(\frac{1}{2}+\epsilon)}P(u^{\alpha_n}\cdot\nabla) u^{\alpha_n}, A^{-\frac{1}{2}-\alpha_n +\epsilon} \phi>_{[L^2(\Omega)]^2} dt   \\
&\to \int_0^T <A^{-(\frac{1}{2}+\epsilon)}P(u\cdot\nabla)u, A^{-\frac{1}{2}+\epsilon} \phi>_{[L^2(\Omega)]^2} dt.  
\end{split}
\end{equation}  
\end{remark}  

The construction presented above allows us to formulate the following theorem:  
\begin{theorem}\label{critth}
Let $\{ \au \}_{\alpha\in (0, \frac{1}{2}]}$ be the set of regular $D(A^{\frac{3}{4}})$ solutions to sub-critical equations \eqref{fractal}. Such solutions are  bounded in the space $[L^2(\Omega)]^2$, uniformly in $\alpha$. As a consequence of that and the regularity properties of such  solutions (varying in $D(A^{\frac{3}{4}}) \subset [H^{\frac{3}{2}}(\Omega)]^2$), for arbitrary sequence $\{ \alpha_n \} \subset (0, \frac{1}{2}]$ convergent to $0^+$ we can find a subsequence $\{ \alpha_{n_k} \}$ that the corresponding sequence of solutions $\{ u^{\alpha_{n_k}} \}$ converges weakly in $[L^2(\Omega)]^2$ to a function $u$ fulfilling the equation:  
\begin{equation}\label{2.11}  
\forall_{\phi\in X_2}  \   \omega_\phi = -<u, \phi>_{[L^2(\Omega)]^2} + <Pf, A^{-1} \phi>_{[L^2(\Omega)]^2}.
\end{equation}  
Due to denseness of the set $D(A^{\frac{3}{4}})$ in $X_2$, the right hand side of \eqref{2.11} defines a unique element in $X_2$. The left hand side $\omega_\phi$ is defined in \eqref{added} and discussed in Remark \ref{explanation}. 
\end{theorem} 

\begin{remark}
As well known (e.g. \cite[Theorem 6.2]{Li}), the $L^\infty(0,T; [L^2(\Omega)]^2) \cap L^2(0,T; D(A^{\frac{1}{2}}))$ solution of the 2-D N-S equation having $u_t \in L^2(0,T; D(A^{-\frac{1}{2}}))$ is unique. Indeed, if $u_1, u_2$ are two such (weak) solutions, applying projector $P$, taking the difference of the equations and multiplying the result by $w = u_1-u_2$, we get:
\begin{equation}\label{bum}
\frac{1}{2} \frac{d}{dt} \| w(t) \|^2_{[L^2(\Omega)]^2} \leq -\| A^{\frac{1}{2}}w \|^2_{[L^2(\Omega)]^2} + <P(w\cdot \nabla)u_1, w>_{[L^2(\Omega)]^2} + <P(u_2\cdot \nabla)w, w>_{[L^2(\Omega)]^2}.
\end{equation}
The last term vanishes for divergence-free functions. The earlier term, using an equivalent form of the nonlinearity in \eqref{equivalent} and the Nirenberg-Gagliardo inequality, is estimated as follows:  
\begin{equation}
\begin{split}
|<P(w\cdot \nabla)u_1, w>_{[L^2(\Omega)]^2}| &\leq c\|w\|^2_{[L^4(\Omega)]^2} \|A^{\frac{1}{2}}u_1\|_{[L^2(\Omega)]^2}   \\ 
&\leq c\|w\|_{[L^2(\Omega)]^2} \|A^{\frac{1}{2}} w\|_{[L^2(\Omega)]^2} \| A^{\frac{1}{2}}u_1\|_{[L^2(\Omega)]^2}. 
\end{split} 
\end{equation}
Inserting the last estimate into \eqref{bum}, using Cauchy's inequality, we obtain 
\begin{equation*}
\begin{split}
\frac{1}{2}\frac{d}{dt} \| w(t) \|^2_{[L^2(\Omega)]^2} &\leq -\| A^{\frac{1}{2}}w \|^2_{[L^2(\Omega)]^2} + c\|w\|_{[L^2(\Omega)]^2} \|A^{\frac{1}{2}}w\|_{[L^2(\Omega)]^2} \|A^{\frac{1}{2}}u_1\|_{[L^2(\Omega)]^2}   \\
&\leq C \| w\|^2_{[L^2(\Omega)]^2} \|A^{\frac{1}{2}}u_1\|^2_{[L^2(\Omega)]^2}. 
\end{split}
\end{equation*} 
Since $\|w(0)\|^2_{[L^2(\Omega)]^2} =0$ then $\| w(t) \|^2_{[L^2(\Omega)]^2}=0$ for all $t \in [0,T]$, due to the Gronwall lemma.    
\end{remark}

\begin{remark}
In 2-D, convergence of the approximated solutions $\au$ of \eqref{fractal} to the unique solution $u$ of the N-S equation holds in a better sense when $u_0\in D(A^{\frac{1}{2}})$. Multiplying \eqref{fractal} by $A\au$, we get 
\begin{equation}
\frac{1}{2} \frac{d}{dt} \| A^{\frac{1}{2}}\au\|_{[L^2(\Omega)]^2}^2 = -\|A^{1+\frac{\alpha}{2}}\au\|_{[L^2(\Omega)]^2}^2  + <P(\au \cdot\nabla)\au, A\au>_{[L^2(\Omega)]^2} + <Pf,A\au>_{[L^2(\Omega)]^2}.
\end{equation}
Further, due to \eqref{2crit}, Nirenberg-Gagliardo and Young inequalities,  
\begin{equation*}
\begin{split}
&|<P(\au \cdot\nabla)\au, A\au>_{[L^2(\Omega)]^2}| \leq \|A^{-\frac{\alpha}{2}}P(\au\cdot\nabla)\au\|_{[L^2(\Omega)]^2} \|A^{1+\frac{\alpha}{2}}\au\|_{[L^2(\Omega)]^2} \\
&\leq c\|\au\|^{1+\alpha}_{[L^4(\Omega)]^2} \|\nabla \au\|^{1-\alpha}_{[L^4(\Omega)]^2} \|A^{1+\frac{\alpha}{2}}\au\|_{[L^2(\Omega)]^2}  \leq \epsilon\|A^{1+\frac{\alpha}{2}}\au\|^2_{[L^2(\Omega)]^2} + C_\epsilon \|\au\|^{\frac{2[1+\alpha- 2\alpha^2]}{1+5\alpha}}_{[H^1(\Omega)]^2} \|\au\|^{\frac{2[2+3\alpha+\alpha^2]}{1+5\alpha}}_{[L^2(\Omega)]^2}. 
\end{split}
\end{equation*}
For small $\epsilon >0$, $\alpha >0$, we thus have  
\begin{equation}
\frac{1}{2} \frac{d}{dt} \| A^{\frac{1}{2}}\au\|_{[L^2(\Omega)]^2}^2 \leq -\frac{1}{4}\|A^{1+\frac{\alpha}{2}}\au\|^2_{[L^2(\Omega)]^2}  + C_\epsilon \|\au\|^{\frac{2[1+\alpha- 2\alpha^2]}{1+5\alpha}}_{[H^1(\Omega)]^2} \|\au\|^{\frac{2[2+3\alpha+\alpha^2]}{1+5\alpha}}_{[L^2(\Omega)]^2}  + \|Pf\|^2_{[L^2(\Omega)]^2}. 
\end{equation}
It follows from the natural a priori estimates that the family $\{ \au \}_{\alpha \in (0, \frac{1}{2}]}$, is bounded in the norm 
\begin{equation}
L^\infty(0,T; [H^1_0(\Omega)]^2) \cap L^2(0,T; D(A)).
\end{equation}
Consequently, it is {\it precompact} in 
\begin{equation}
L^p(0,T; [H^{1^-}_0(\Omega)]^2) \cap L^2(0,T; D(A^{1^-})),
\end{equation}
with arbitrary $p>1$. The convergence $\au \to u$ as $\alpha \to 0^+$ is thus verified in the above space.     
\end{remark}

\begin{remark}
To understand better the utility of the method used in the study of the global extendibility of solutions we will discuss shortly an example\footnote{suggested kindly by professor Eduard Feireisl} of the {\em Burgers type system in 3-D}, obtained by neglecting the viscosity in homogeneous 3-D N-S equation:
\begin{equation}\label{Burgers}
\begin{split}
&U_t = \nu \Delta U - (U \cdot\nabla)U,   \quad x \in \Omega,\, t>0,  \\
&U = 0, \quad t>0, \quad x \in \partial \Omega,   \\ 
&U(0,x) = U_0(x).
\end{split}
\end{equation}
It is easy to see that each component of the sufficiently regular (that is, varying in $[L^\infty(\Omega)]^3$) solution of \eqref{Burgers} fulfills Maximum Principle:
\begin{equation}
\| U_i(t,\cdot)\|_{L^\infty(\Omega)} \leq \| U_{0i} \|_{L^\infty(\Omega)}, \  i=1,2,3.  
\end{equation}
We are thus given a natural a priori estimate in $Y=[L^\infty(\Omega)]^3$ for such system. The nonlinear term is as in the N-S equation, and we have the estimates:  
\begin{equation}\label{burgest}
\begin{split}
&\|(U\cdot \nabla)U\|_{[L^2(\Omega)]^3} \leq \| U \|_{[L^\infty(\Omega)]^3} \| U \|_{[H^1_0(\Omega)]^3} \leq \| U_0 \|_{[L^\infty(\Omega)]^3} \| U \|_{[H^1_0(\Omega)]^3},  \\
&\|(U\cdot \nabla)U\|_{[H^{-1}(\Omega)]^3} \leq c \| U \|^2_{[L^\infty(\Omega)]^3},
\end{split}
\end{equation}
valid in particular for all the local solutions varying in $[H^{\frac{3}{2}^+}(\Omega)]^3 \cap [H^1_0(\Omega)]^3$ (as in Theorem \ref{localexistence}). Since $H^{\frac{3}{2}^+}(\Omega)\subset L^\infty(\Omega)$, the Maximum Principle works for such solutions.  By interpolation, \eqref{burgest} gives   
\begin{equation}
\|(U\cdot \nabla)U\|_{[H^{-\frac{1}{4}}(\Omega)]^3} \leq c  \| U \|^{\frac{5}{4}}_{[L^\infty(\Omega)]^3} \|U \|^{\frac{3}{4}}_{[H^1_0(\Omega)]^3} \leq c \| U_0 \|^{\frac{5}{4}}_{[L^\infty(\Omega)]^3} \|U \|^{\frac{3}{4}}_{[H^1_0(\Omega)]^3},
\end{equation}
which shows the nonlinearity is {\em sub-critical} in that case. Consequently, the local solution will be extended globally in time.   

The above example shows that our approach is sensitive not only on the form of nonlinearity, but also another specific properties. It also indicates the role of {\em the pressure} in the classical N-S equation, the term which seems responsible for the delicate properties of solutions of that equation.    
\end{remark}

\begin{remark}
It is evident from the considerations above that the phenomenon of {\em loosing regularity by local smooth solutions of 3-D N-S} is possible only if they enter the super-critical range of the nonlinearity (compare \eqref{1.26}). It would be therefore interesting to consider local solutions corresponding to initial data $u_0$ fulfilling 
\begin{equation}
\| A^{-\frac{1}{8}} F(u_0)\|_{[L^2(\Omega)]^3} \geq g(\| u_0\|_{[L^2(\Omega)]^3}) \| A^{\frac{7}{8}}u_0\|^{1+\epsilon}_{[L^2(\Omega)]^3},
\end{equation}
where $0< \epsilon< \frac{2}{7}$, as eventual candidates for such phenomenon.  
\end{remark}

\section{Appendix. Fractional powers operators and estimates.}\label{4.5}  
\subsection{Some technicalities.} When passing to the limit in the considerations above it was important that the estimates, in particular the constants in it, can be taken {\em uniform} in $\alpha$. Therefore, in the technical lemmas below we need to care on a very precise expression of that uniformity. Even some estimates can be found in the literature, usually such uniformity is not clear from the presentation, thus we include here the proofs for completeness.

First, we formulate a lemma used in the previous calculations: 
\begin{lemma}\label{new} 
Let $A$ be a {\em positive operator} in a Banach space $X$ (\cite{TR, M-S, C-D}). For arbitrary  $\phi \in X$, we have  
\begin{equation*}
\forall_{\epsilon >0} \exists_{L}  \|(I - A^{-\beta})\phi \|_{X} \leq \sin(\pi\beta) \bigl(\frac{2L(1+M)}{\pi} + L^{-1}M)\| \phi\|_X + \epsilon.    
\end{equation*}
Consequently, the left hand side tends to zero as $0 < \beta \to 0^+$.    
\end{lemma}
\begin{proof}
Our task is, for fixed $\phi \in X$ and $\beta$ near $0^+$, to estimate the expression:  
\begin{equation}\label{integr}
\begin{split}
(A^{-\beta}-I)\phi &= \frac{\sin(\pi\beta)}{\pi} \int_0^\infty \lambda^{-\beta} (\lambda+A)^{-1} \phi d\lambda - \frac{\sin(\pi(1-\beta))}{\pi} \int_0^\infty \frac{\lambda^{(1-\beta)-1}}{\lambda+1} d\lambda \  \phi  \\   
&= \frac{\sin(\pi\beta)}{\pi} \int_0^\infty \lambda^{-\beta} [(\lambda+A)^{-1} \phi - \frac{1}{\lambda+1} \phi] d\lambda.
\end{split}
\end{equation}
In the estimates we are using the following properties; taken from \cite[p. 62]{M-S} equality valid for $\eta \in (0,1)$   
\begin{equation*}
\int_0^\infty \frac{\lambda^{\eta-1}}{\lambda+1} d\lambda = \frac{\pi}{\sin(\pi\eta)}, 
\end{equation*}
the simple formula:    
\begin{equation}\label{raz}
(\lambda+A)^{-1} \phi - \frac{1}{\lambda+1} \phi = \frac{1}{\lambda+1} \bigl[\lambda(\lambda+A)^{-1} \phi-\phi +(\lambda+A)^{-1} \phi\bigr],      
\end{equation}
and the two asymptotic properties of {\it non-negative operators} valid on functions $\phi \in X$ taken from \cite[Proposition 1.1.3]{M-S}:
\begin{equation}\label{dwaa}
\begin{split}
&\lim_{\lambda \to \infty} \lambda(\lambda+A)^{-1} \phi = \phi,     \\
&\lim_{\lambda\to \infty} (\lambda+A)^{-1} A\phi = 0.
\end{split}  
\end{equation}  

Returning to the proof, we split the integral in \eqref{integr} into $(0,L)$ and $(L,\infty)$ and estimate the first part,  
\begin{equation}
\begin{split}
\frac{\sin(\pi\beta)}{\pi}\| \int_0^L \lambda^{-\beta} (\frac{1}{\lambda+1} - (\lambda+A)^{-1})\phi d\lambda\|_X &\leq \frac{\sin(\pi\beta)}{\pi} \int_0^L \lambda^{-\beta}(1 + M) d\lambda \| \phi\|_X    \\
&=\frac{\sin(\pi\beta)}{\pi} \frac{L^{1-\beta}}{1-\beta} (1+M) \|\phi\|_X,
\end{split}
\end{equation}
where $L>0$ will be chosen later. Note that letting $\beta \to 0^+$ the result of the estimate above is bounded by $|\frac{\sin(\pi\beta)}{\pi}| 2L(1+M)\|\phi\|_X \to 0$, for any fixed $L>0$. 

Next using \eqref{raz}, the integral over $(L, \infty)$ is, for $\phi \in X$, estimated as follows: 
\begin{equation}
\frac{\sin(\pi\beta)}{\pi} \|\int_L^\infty \lambda^{-\beta} [(\lambda+A)^{-1} \phi - \frac{1}{\lambda+1} \phi] d\lambda \|_X \leq \frac{\sin(\pi\beta)}{\pi} \int_L^\infty \frac{\lambda^{-\beta}}{\lambda+1} \| \lambda(\lambda+A)^{-1} \phi-\phi +(\lambda+A)^{-1} \phi \|_X d\lambda, 
\end{equation}
where due to \eqref{dwaa} we see that       
\begin{equation}\label{5.14}  
\|(\lambda+1)(\lambda+A)^{-1} A\phi - \phi\|_X \leq  \| \lambda(\lambda+A)^{-1} \phi-\phi\|_X + \|(\lambda+A)^{-1} \phi \|_X \leq \epsilon +\frac{M}{1+\lambda} \|\phi\|_X  \  \text{as}  \  \lambda \to \infty,
\end{equation}
$\epsilon>0$ arbitrary fixed.  Consequently we obtain:  
\begin{equation}
\begin{split}
\frac{\sin(\pi\beta)}{\pi}  \int_L^\infty \frac{\lambda^{-\beta}}{\lambda +1} \bigl(\epsilon +\frac{M}{1+\lambda} \|\phi\|_X) d\lambda &\leq \frac{\sin(\pi\beta)}{\pi}  \int_L^\infty \lambda^{-\beta-1} \bigl(\epsilon +\frac{M}{\lambda} \|\phi\|_X) d\lambda    \\
&\leq \frac{\sin(\pi\beta)}{\pi} \frac{L^{-\beta}}{\beta}\epsilon + \frac{\sin(\pi\beta)}{\pi} \frac{L^{-\beta-1}}{\beta +1} M \|\phi\|_X,   
\end{split}
\end{equation}    
for sufficiently large value of $L \geq 1$, as specified in \eqref{5.14}. Note that letting $\beta \to 0^+$ in the resulting estimate we have:
\begin{equation}
\frac{\sin(\pi\beta)}{\pi\beta} L^{-\beta} \epsilon + \frac{\sin(\pi\beta)}{\pi} \frac{L^{-\beta-1}}{\beta +1} M \|\phi\|_X \leq \epsilon + \sin(\pi\beta) L^{-1} M \|\phi\|_X,
\end{equation}
for chosen large value of $L$.   

For such $L$ we get a final estimate of the integral in \eqref{integr} having the form:     
\begin{equation}\label{5.17}
\| (A^{-\beta}-I)\phi\|_X \leq \frac{\sin(\pi\beta)}{\pi} \|\int_0^\infty \lambda^{-\beta} [(\lambda+A)^{-1} \phi - \frac{1}{\lambda+1} \phi] d\lambda \|_X \leq \sin(\pi\beta) \bigl(\frac{2L(1+M)}{ \pi} + L^{-1} M \bigr)\|\phi\|_X + \epsilon,    
\end{equation}
where $\epsilon >0$ was arbitrary. The right hand side of \eqref{5.17} will be made small when we let $\beta$ near $0^+$, noting $\epsilon$ was an arbitrary positive number.   
\end{proof}

\subsection{Properties of the fractional powers operators.}
Recall first the Balakrishnan definition of fractional power of non-negative operator (e.g. \cite[p. 299]{KO}).  Let $A$ be a closed linear densely defined operator in a  Banach space $X$, such that its resolvent set contains $(-\infty,0)$ and the resolvent satisfies:
\begin{equation*}
\| \lambda(\lambda+A)^{-1}\| \leq M, \ \lambda>0.
\end{equation*}
Then, for $\eta \in (0,1)$,
\begin{equation}\label{flaps}
A^\eta \phi = \frac{\sin(\pi\eta)}{\pi} \int_0^\infty s^{\eta - 1} A (s + A)^{-1} \phi ds.
\end{equation}
Note that there is another definition, through singular integrals, of the fractional powers of the $(-\Delta)^{-\alpha}$ in $L^p(\R^N)$ frequently used in the literature. See \cite[Chapter 2.2]{M-S} for the proof of equivalence of the two definitions for $1<p<\frac{N}{2\Re \alpha}$; see also \cite[section 4.3]{D-K-S1}.  

{\bf Moment inequality.} We recall here, the {\it moment inequality} estimate valid for fractional powers of non-negative operators. They are suitable to compare various fractional powers. Recalling \cite[p. 98]{Ya}, we have the following version of the moment inequality with precise constant; for $0 \leq \alpha < \beta < \gamma \leq 1$, 
\begin{equation*}
\| A^\beta \phi\| \leq \frac{(\sin \frac{\pi(\beta-\alpha)}{\gamma -\alpha}) (\gamma -\alpha)^2}{\pi(\gamma -\beta)(\beta -\alpha)} (M +1) \| A^\gamma \phi \|^{\frac{\beta -\alpha}{\gamma -\alpha}} \| A^\alpha \phi \|^{\frac{\gamma -\beta}{\gamma -\alpha}},
\end{equation*}
where $\phi \in D(A^\gamma)$. 

\subsection{Moment inequality extended.} The task here is to extend the moment inequality to the form suitable to compare the powers $1+\alpha$ and $1$  (here $\alpha>0$ near $0^+$). We need to use a more general than \eqref{flaps} expression \eqref{generalexpression} for the fractional powers, taken from \cite[(3.4), p.59]{M-S}, which states that:  
\begin{equation}
A^{1+\alpha} \phi = \frac{2 \sin(\pi\alpha)}{(1-\alpha)\pi} \int_0^\infty \lambda^\alpha [A(\lambda+A)^{-1}]^2 \phi d\lambda,  \  \phi\in D(A^2), 
\end{equation}
where the original term $\frac{\Gamma(2)}{\Gamma(\alpha) \Gamma(2-\alpha)}$ has been transformed using the known properties of the $\Gamma$ function; 
$$
\Gamma(1+\alpha) = \alpha \Gamma(\alpha),\quad \Gamma(\alpha) \Gamma(1-\alpha) = \frac{\pi}{\sin(\pi \alpha)},\quad \Gamma(2)  = 2.
$$ 
We are using the following bound, valid for {\it positive operators}, in the calculations below:
\begin{equation}
\|[A(\lambda +A]^{-1}\|_X \leq M+1,
\end{equation}
splitting the integral over $(0,L)$ and $(L,\infty)$, we estimate the first integral,   
\begin{equation}
\begin{split}
\frac{2\sin(\pi\alpha)}{(1-\alpha)\pi} \int_0^L \lambda^\alpha [A(\lambda+A)^{-1}]^2 \phi d\lambda &= \frac{2 \sin(\pi\alpha)}{(1-\alpha)\pi} \int_0^L \lambda^\alpha (\lambda +A)^{-1} [A(\lambda+A)^{-1}] A\phi d\lambda    \\
&= \frac{2\sin(\pi\alpha)}{(1-\alpha)\pi} \int_0^L \lambda^\alpha \frac{M}{1+\lambda} (M+1) d\lambda \| A\phi \|_X. 
\end{split}    
\end{equation}
We thus have:
\begin{equation}
\|\frac{2 \sin(\pi\alpha)}{(1-\alpha)\pi} \int_0^L \lambda^\alpha [A(\lambda+A)^{-1}]^2 \phi d\lambda \|_X \leq \frac{2\sin(\pi\alpha)}{(1-\alpha)\pi} M(M+1) \frac{L^\alpha}{ \alpha} \|A\phi\|_X. 
\end{equation}
  
The second integral over $(L,\infty)$ is estimated next:  
\begin{equation}
\begin{split}
\frac{2\sin(\pi\alpha)}{(1-\alpha)\pi} \|\int_L^\infty \lambda^\alpha [(\lambda +A)^{-1}]^2 A^2\phi d\lambda \|_X &\leq  \frac{2\sin(\pi\alpha)}{(1-\alpha)\pi} \int_L^\infty \lambda^\alpha \bigl(\frac{M}{ \lambda}\bigr)^2 \|A^2\phi\|_X d\lambda     \\
&= \frac{2\sin(\pi\alpha)}{(1-\alpha)\pi} \frac{M^2}{1-\alpha} L^{\alpha-1} \|A^2\phi\|_X. 
\end{split}
\end{equation}
Minimizing with respect to $L>0$ we get $L_{min} = \frac{M\|A^2\phi\|_X}{(M+1)\|A\phi\|_X}$, which leads to the final estimate:
\begin{equation}
\begin{split}
\| A^{1+\alpha} \phi\|_X &\leq \frac{2\sin(\pi\alpha)}{(1-\alpha)\pi} M(M+1) \frac{L_{min}^\alpha}{ \alpha} \|A\phi\|_X + \frac{2\sin(\pi\alpha)}{(1-\alpha)\pi} \frac{M^2}{1-\alpha} L_{min}^{\alpha-1} \|A^2\phi\|_X    \\
& =\frac{2\sin(\pi\alpha)}{(1-\alpha)^2 \pi\alpha} M^{1+\alpha} (M+1)^\alpha \| A\phi\|_X^{1-\alpha} \| A^2 \phi\|_X^\alpha \to 2M \| A\phi\|_X,        
\end{split} 
\end{equation}  
as $\alpha \to 0^+$. 

\section{Conclusion.}  
The PROBLEM connected with the 3-D Navier-Stokes equation is not solved in the paper. But we are explaining its reason. The viscosity in the classical N-S equation is too weak to prevent better norms of its solutions from blowing-up in a finite time, especially in 3-D case. The 2-D equation is still {\em critical} with respect to the standard $L^2$ estimates. In order, for small initial data we enjoy the standard property of problems with {\em quadratic nonlinearity}, that near zero function the estimates of solutions obtained from the equation are proportional to the square of the norm, therefore there is a ball centered at zero the solutions originated in it will never leave that ball. For such solutions, the nonlinear term in N-S is {\em subordinated} to the main part operator. Concluding; {\em to solve the problem with the N-S one need to find a better a priori estimate than the standard one in $L^2$, coming from the 'symmetry' of that equation. Then, the nonlinear term would be subordinated to the main part operator preventing possible blow-up of better norms of the solutions, and the global in time existence-uniqueness theory will apply.} In particular, if the estimate in $L^N(\Omega)$ ($L^{N^+}(\Omega)$) is known, it will make the N-D Navier-Stokes equation critical (sub-critical) due to the following calculation
\begin{equation}   
\|P(u \cdot \nabla)u\|_{X^{-\frac{1}{2}}} \leq c\| |u|^2\|_{[L^2(\Omega)]^N} \leq c' \| u \|_{[L^{\frac{2N}{N-2}}(\Omega)]^N} \| u\| _{[L^N(\Omega)]^N}  
\leq c'' \| u \|_{X^{\frac{1}{2}}} \| u\| _{[L^N(\Omega)]^N}.
\end{equation}

It seems that the fractal regularizations; \eqref{fractal} in 2-D and \eqref{reg}, $s>\frac{5}{4}$, in 3-D, having much better properties of solutions, offer an alternative way of description of the flow. But until very recently nobody was using fractal equations as models of physical phenomena.

\bigskip
{\bf Acknowledgement.} The paper is dedicated to my mentors, who teach me through the years 1980-90. 

The author is supported by NCN grant DEC-2012/05/B/ST1/00546 (Poland).


\begin{thebibliography}{11}
\bibitem{AM} H. Amann, \textit{Linear and Quasilinear Parabolic Problems, Volume I, Abstract Linear Theory}, Birkha\"{u}ser Verlag, Basel, 1995.  

\bibitem{C-D1} J.W. Cholewa, T. Dlotko, \textit{Local attractor for $n-D$ Navier-Stokes system}, Hiroshima Math. J. 28 (1998), 309-319.  

\bibitem{C-D} J.W. Cholewa, T. Dlotko, \textit{Global Attractors in Abstract Parabolic Problems}, Cambridge University Press, Cambridge, 2000.

\bibitem{C-C-W} P. Constantin, D. Cordoba, J. Wu, \textit{On the critical dissipative Quasi-geostrophic equation}, Indiana  Univ. Math. J. 50 (2001), 97-108.

\bibitem{C-C} A. Cordoba, D. Cordoba, \textit{A Maximum Principle applied to quasi-geostrophic equations}, Commun. Math. Phys. 249 (2004), 511-528.

\bibitem{C-C1} A. Cordoba, D. Cordoba, \textit{A pointwise estimate for fractionary derivatives with applications to partial differential equations},
Proc. Natl. Acad. Sci. USA 100 (2003), 15316-15317.

\bibitem{DL} T. Dlotko, \textit{The generalized Korteweg-de Vries-Burgers equation in $H^2(\R)$}, Nonlinear Anal. TMA 74 (2011), 721-732.

\bibitem{D-S} T. Dlotko, C. Sun, \textit{Dynamics of the modified viscous Cahn-Hilliard equation in $\mathbb{R}^N$}, Topol. Methods Nonlinear Anal. 35 (2010), 277-294.

\bibitem{D-K-S} T. Dlotko, M.B. Kania, C. Sun, \textit{Analysis of the viscous Cahn-Hilliard equation in $\R^N$}, J. Differential Equations 252 (2012), 2771-2791. 

\bibitem{D-K} T. Dlotko, M.B. Kania, \textit{Subcritical Hamilton-Jacobi fractional equation in $\mathbb{R}^N$}, Math. Methods Appl. Sci. (2015), DOI:10.1002/mma.3241.  

\bibitem{D-K-S1} T. Dlotko, M.B. Kania, C. Sun, \textit{Quasi-geostrophic equation in $R^2$}, J. Differential Equations 259 (2015), 531-561. 

\bibitem{D-K-Sh} T. Dlotko, M.B. Kania, Shan Ma, \textit{Korteweg-de Vries-Burgers system in $\R^N$}, J. Math. Anal. Appl. 411 (2014), 853-872. 

\bibitem{D-K-Y}  T. Dlotko, M. B. Kania, Meihua Yang, \textit{Generalized Korteweg-de Vries equation in $H^1(R)$}, Nonlinear Analysis TMA 71 (2009), 3934-3947.    

\bibitem{F-H-T} C. Foias, D.D. Holm, E.S. Titi,  \textit{The Navier–-Stokes--alpha model of fluid turbulence}, Physica D: Nonlinear Phenomena, 2001.  

\bibitem{F-T} C. Foias, O. Manley, R. Rosa, R. Temam, \textit{Navier–-Stokes Equations and Turbulence}, Encyclopedia of Mathematics and its Applications, Cambridge University Press, Cambridge, 2004. 

\bibitem{FR} A. Friedman, \textit{Partial Differential Equations}, Holt, Rinehart and Winston, New York, 1969.  

\bibitem{GI} Y. Giga, \textit{Analyticity of the semigroup generated by the Stokes operator in $L_r$ spaces}, Math. Z. 178 (1981), 297-329. 

\bibitem{GI1} Y. Giga, \textit{Domains of fractional powers of the Stokes operator in $L_r$ spaces}, Arch. Rational Mech. Anal. 89 (1985), 251-265.  

\bibitem{G-M} Y. Giga, T. Miyakawa, \textit{Solutions in $L_r$ of the Navier-Stokes initial value problem}, Arch. Rational Mech. Anal. 89 (1985), 267-281.  

\bibitem{HE} D. Henry, \textit{Geometric Theory of Semilinear Parabolic Equations}, Springer-Verlag, Berlin, 1981. 

\bibitem{HE1} D. Henry, \textit{How to remember the Sobolev inequalities}, in: Differential Equations, Eds. D.G. de Figueiredo and C.S. H\"onig, Springer-Verlag, New York, 1982, pp. 97-109. 

\bibitem{K-N-V} A. Kiselev, F. Nazarov, A. Volberg, \textit{Global well-posedness for the critical $2D$ dissipative quasi-geostrophic equation}, Invent. Math. 167
(2007), 445-453.

\bibitem{KO} H. Komatsu, \textit{Fractional powers of operators}, Pacific J. Math. 19 (1966), 285-346.

\bibitem{KO1} H. Komatsu, \textit{Fractional powers of operators, II. Interpolation spaces}, Pacific J. Math. 21 (1967), 89-111. 

\bibitem{L} J. Leray, \textit{Essai sur le mouvement d'un fluide visqueux emplissant l'espace}, Acta Math. 63 (1934), 193-248.

\bibitem{L-P} F. Linares, G. Ponce, \textit{Introduction to Nonlinear Dispersive Equations}, IMPA, Rio de Janeiro, 2008.

\bibitem{Li} J.-L. Lions, \textit{Quelques m\'ethodes de r\'esolution des probl$\grave{e}$mes aux limites non lin\'eaires}, Dunod Gauthier-Villars, Paris, 1969.

\bibitem{M-S} C. Mart\'inez Carracedo, M. Sanz Alix, \textit{The Theory of Fractional Powers of Operators}, Elsevier, Amsterdam, 2001.

\bibitem{BE} A. Rodriguez-Bernal, \textit{Existence, uniqueness and regularity of solutions of nonlinear evolution equations in extended scales of Hilbert spaces}, 
CDSNS91-61 Report, Georgia Institute of Technology, Atlanta, 1991. 

\bibitem{Sz} J. Szarski, \textit{Differential Inequalities}, PWN--Polish Scientific Publishers, Warszawa, 1967.  

\bibitem{TE1} R. Temam, \textit{Navier-Stokes Equations, Theory and Numerical Analysis}, North-Holland, Amsterdam, 1979. 

\bibitem{TE2} R. Temam, \textit{On the Euler equations of incompressible perfect fluids}, J. Functional Analysis 20 (1975), 32-43.  

\bibitem{TR} H. Triebel, \textit{Interpolation Theory, Function Spaces, Differential Operators}, Veb Deutscher Verlag, Berlin, 1978.

\bibitem{WA} W. von Wahl, \textit{Equations of Navier-Stokes and Abstract Parabolic Equations}, Vieweg, Braunschweig/Wiesbaden, 1985. 

\bibitem{Wa} W. Walter, \textit{Differential and Integral Inequalities}, Springer-Verlag, New York, 1970.  

\bibitem{W1} J. Wu, \textit{Dissipative quasi-geostrophic equations with $L^p$ data},  Electron. J. Differential Equations 56 (2001), 1-13.

\bibitem{W2} J. Wu, \textit{The quasi-geostrophic equation and its two regularizations}, Comm. Partial Differ. Equ. 27 (2002), 1161-1181.

\bibitem{Ya} A. Yagi, \textit{Abstract Parabolic Evolution Equations and their Applications}, Springer-Verlag, Heidelberg, 2010. 
\end{thebibliography}
\end{document}